\documentclass[11pt]{article}
\usepackage[T1]{fontenc}      
\usepackage{latexsym}
\usepackage{amssymb}
\usepackage{amsmath}
\usepackage{graphicx}
\usepackage{colortbl}
\usepackage{tikz}
\usetikzlibrary{decorations.pathreplacing}
\usetikzlibrary{patterns}
\usepackage{nicefrac}
\usepackage{subfigure}
\usepackage{todonotes}
\usepackage[square,comma]{natbib} 

\begin{document}

\def\Nset{\mathbb{N}}
\def\Ascr{\mathcal{A}}
\def\Bscr{\mathcal{B}}
\def\Cscr{\mathcal{C}}
\def\Dscr{\mathcal{D}}
\def\Escr{\mathcal{E}}
\def\Fscr{\mathcal{F}}
\def\Hscr{\mathcal{H}}
\def\Iscr{\mathcal{I}}
\def\Mscr{\mathcal{M}}
\def\Nscr{\mathcal{N}}
\def\Pscr{\mathcal{P}}
\def\Cscr{\mathcal{C}}
\def\Rscr{\mathcal{R}}
\def\Sscr{\mathcal{S}}
\def\Uscr{\mathcal{U}}
\def\Wscr{\mathcal{W}}
\def\Xscr{\mathcal{X}}
\def\cupp{\stackrel{.}{\cup}}
\def\stern{\textasteriskcentered}
\def\bold{\bf\boldmath}
\def\tp{{\scriptscriptstyle\top}}
\def\Ssumeven{\sum_{S\in\Sscr \;\! : \;\! |S\cap C| \,\text{even}}}
\def\Ssumone{\sum_{S\in\Sscr \;\! : \;\! |S\cap C|=1}}
\def\Csumeven{\sum_{C\in\Cscr \;\! : \;\! |S\cap C| \,\text{even}}}
\def\Csumone{\sum_{C\in\Cscr \;\! : \;\! |S\cap C|=1}}
\def\bomc{\text{\rm BOMC}}

\newcommand{\rouge}[1]{\textcolor{red}{\tt \footnotesize #1}}

\newcommand{\boldheader}[1]{\smallskip\noindent{\bold #1:}\quad}
\newcommand{\PP}{\mbox{\slshape P}}
\newcommand{\NP}{\mbox{\slshape NP}}
\newcommand{\opt}{\mbox{\scriptsize\rm OPT}}
\newcommand{\ec}{\mbox{\scriptsize\rm OPT}_{\small\rm 2EC}}
\newcommand{\lp}{\mbox{\scriptsize\rm LP}}
\newcommand{\inn}{\mbox{\rm in}}
\newcommand{\deff}{\mbox{\rm sur}}
\newcommand{\MAXSNP}{\mbox{\slshape MAXSNP}}
\newtheorem{theorem}{Theorem}
\newtheorem{lemma}[theorem]{Lemma}
\newtheorem{corollary}[theorem]{Corollary}
\newtheorem{proposition}[theorem]{Proposition}
\newtheorem{definition}[theorem]{Definition}
\def\prove{\par \noindent \hbox{\bf Proof:}\quad}
\def\endproof{\eol \rightline{$\Box$} \par}
\renewcommand{\endproof}{\hspace*{\fill} {\boldmath $\Box$} \par \vskip0.5em}
\newcommand{\mathendproof}{\vskip-1.8em\hspace*{\fill} {\boldmath $\Box$} \par \vskip1.8em}
\def\cupp{\stackrel{.}{\cup}}
\newcommand{\sfrac}[2]{\textstyle{\frac{#1}{#2}}}

\newcommand{\citegenitiv}[1]{\citeauthor{#1}' [\citeyear{#1}]}
\newcommand{\citegenitivs}[1]{\citeauthor{#1}'s [\citeyear{#1}]}

\definecolor{orange}{rgb}{1,0.5,0}
\definecolor{violet}{rgb}{0.8,0,1}
\definecolor{darkgreen}{rgb}{0,0.5,0}
\definecolor{grey}{rgb}{0.6,0.6,0.6}
\definecolor{lightgrey}{rgb}{0.7,0.7,0.7}
\definecolor{turq}{rgb}{0,0.4,0.4}
\definecolor{darkgreen}{rgb}{0.1, 0.6, 0.3}

\newcommand{\peven}{p^C_{\text{\rm even}}}
\newcommand{\pone}{p^C_{\text{\rm one}}}
\newcommand{\pmany}{p^C_{\text{\rm many}}}

\setcounter{topnumber}{9}
\setcounter{bottomnumber}{9}
\setcounter{totalnumber}{9}
\renewcommand{\topfraction}{0.99}
\renewcommand{\bottomfraction}{0.99}
\renewcommand{\textfraction}{0.01}

\title {
\vspace*{-1.8cm}
{\bf\boldmath Better $s$-$t$-Tours by Gao Trees} 
}
\author{\Large Corinna Gottschalk \qquad Jens Vygen \\[1mm] \small RWTH Aachen \qquad University of Bonn}

\date{\small November 17, 2015\thanks{This work was done during the trimester program on combinatorial optimization
at the Hausdorff Institute for Mathematics in Bonn.}}

\begingroup
\makeatletter
\let\@fnsymbol\@arabic
\maketitle
\endgroup

\parindent0pt

\begin{abstract}
We consider the $s$-$t$-path TSP: given a finite metric space with two elements $s$ and $t$, 
we look for a path from $s$ to $t$ that contains all the elements and has minimum total distance.
We improve the approximation ratio for this problem from 1.599 to 1.566.
Like previous algorithms, we solve the natural LP relaxation and represent an optimum solution $x^*$
as a convex combination of spanning trees.
Gao showed that there exists a spanning tree in the support of $x^*$ that has only one edge in each narrow cut
(i.e., each cut $C$ with $x^*(C)<2$).
Our main theorem says that the spanning trees in the convex combination can be chosen such that
many of them are such ``Gao trees''.

\medskip\noindent
\noindent{{\bf keywords:} traveling salesman problem, $s$-$t$-path TSP, approximation algorithm, spanning tree}
\end{abstract}

\section{Introduction}\label{sec:Introduction}
The traveling salesman problem (TSP) is one of the best-known NP-hard problems in combinatorial optimization. 
In this paper, we consider the $s$-$t$-path variant: Given a finite set $V$ (of \emph{cities}), 
two elements $s,t\in V$ and $c: V\times V \rightarrow \mathbb{R}_{\ge 0}$,  
the goal is to find a sequence $v_1, \ldots, v_N$ containing all cities and with $v_1 = s$ and $v_N = t$, 
minimizing $\sum_{i = 1}^{N - 1} c(v_i, v_{i + 1})$. 
An equivalent formulation requires that every city is visited exactly once, but then we need to assume that $c$ satisfies the triangle inequality.
We work with the latter formulation (to obtain such an instance from a general function $c$, compute the metric closure).
The special case where $s=t$ is the well-known metric TSP; but in this paper we assume $s\not=t$.

The classical algorithm by \cite{Chr76} computes a minimum-cost spanning tree $(V,S)$ and then does 
\emph{parity correction} by adding a minimum-cost
matching on the vertices whose degree in $S$ has the wrong parity.
While Christofides' algorithm is still the best known approximation algorithm for metric TSP (with ratio $\frac{3}{2}$), 
there have recently been improvements for special cases and variants (see e.g.\ \cite{Vyg12} for a survey), including the $s$-$t$-path TSP.

\subsection{Previous Work}

For the $s$-$t$-path TSP, 
Christofides'  algorithm has only an approximation ratio of $\frac{5}{3}$ as shown by \cite{Hoo91}.  
\cite{AnKS12} were the first to improve on this and obtained an approximation ratio of $\frac{1+\sqrt{5}}{2} \approx 1.618$.
They first solve the natural LP relaxation (essentially proposed by \cite{DanFJ54}) and represent
an optimum solution $x^ *$ as a convex combination of spanning trees. 
This idea, first proposed by \cite{HelK70}, was exploited earlier for different TSP variants by \cite{AsaXX10} and \cite{OveSS11}.
Given this convex combination, \cite{AnKS12} do parity correction for each of the contributing trees and output the best of these solutions.
\cite{Seb13} improved the analysis of this \emph{best-of-many Christofides algorithm} and obtained the approximation ratio $\frac{8}{5}$.
\cite{Gao15} gave a unified analysis.
\cite{Vyg15} suggested to ``reassemble'' the trees: starting with an arbitrary convex combination of spanning trees,
he computed a different one, still representing $x^ *$, that avoids certain bad local configurations. This led to the slightly
better approximation ratio of $1.599$.

In this paper, we will reassemble the trees more systematically to obtain a convex combination with strong global properties.
This will enable us to control the cost of parity correction much better, leading to an approximation ratio of $1.566$.

This also proves an upper bound of $1.566$ on the integrality ratio of the natural LP.
The only known lower bound is $1.5$. \cite{SebV12} proved that the integrality ratio is indeed 1.5 for the
graph $s$-$t$-path TSP, i.e., the
special case of \emph{graph metrics} (where $c(v,w)$ is the distance from $v$ to $w$ in a given unweighted graph on vertex set $V$).
\cite{Gao13} gave a simpler proof of this result, which inspired our work: see Section \ref{subsection:gao}.

\subsection{Notation and Preliminaries}

Throughout this paper, $V$ denotes the given set of cities, $n:=|V|$, and $E$ denotes the set of edges
of the complete graph on $V$. Moreover, $c: V\times V \to \mathbb{R}_{\ge 0}$ is the given metric.
For any $U\subseteq V$ we write $E[U]$ for the set of edges with both endpoints in $U$ and $\delta(U)$ for the set of edges with
exactly one endpoint in $U$; moreover, $\delta(v):=\delta(\{v\})$ for $v\in V$. 
If $F\subseteq E$ and $U\subseteq V$, we denote by $(V,F)[U]$ the subgraph $(U,F\cap E[U])$ of $(V,F)$ induced by $U$.
For $x\in\mathbb{R}_{\ge 0}^E$ we write $c(x):=\sum_{e=\{v,w\}\in E}c(v,w)x_e$ and $x(F):=\sum_{e\in F}x_e$ for $F\subseteq E$.
Furthermore, $\chi^{F} \in \{0, 1\}^E$ denotes the characteristic vector of a subset $F \subseteq E$, and $c(F):= c(\chi^F)$.  
For $F\subset E$ and $f \in E$, we write $F + f$ and $F-f$ for $F\cup \{f\}$ and $F\setminus \{f\}$, respectively. 

For $T\subseteq V$ with $|T|$ even, a \emph{$T$-join} is a set $J \subseteq E$ for which $|\delta(v)\cap J|$ is odd if and only if $v \in T$.  \cite{Edm65} proved that a minimum weight $T$-join can be computed in polynomial time. Moreover, the minimum cost of a $T$-join is the minimum over $c(y)$ for $y$ in the 
\emph{$T$-join polyhedron} $\{ y \in \mathbb{R}_{\ge 0}^E: y(\delta(U)) \ge 1 \ \forall U \subset V \text{ with } |U \cap T| \text{ odd} \}$ as proved by \cite{EdmJ73}.
We are going to use that the cost of any vector in the $T$-join polyhedron is an upper bound for the minimum cost of  a $T$-join. 

To obtain a solution for the $s$-$t$-path TSP, 
it is sufficient to compute a connected multi-graph with vertex set $V$ in which exactly $s$ and $t$ have odd degree. 
We call such a graph an \emph{$\{s, t\}$-tour}.
 As an $\{s, t\}$-tour contains an Eulerian walk from $s$ to $t$, we can obtain a \emph{Hamiltonian} $s$-$t$-path (i.e.\ an $s$-$t$-path with vertex set $V$) by traversing the Eulerian walk and shortcutting when the walk encounters a vertex that has been visited already. 
Since $c$ is a metric, it obeys the triangle inequality. Thus, the resulting path is not more expensive than the multigraph. 

By $\Sscr$ we denote the set of edge sets of spanning trees in $(V,E)$.
For $S\in\Sscr$, $T_S$ denotes the set of vertices whose degree has the wrong parity, 
i.e., even for $s$ or $t$ and odd for $v \in V\setminus \{s, t\}$.  
Christofides' algorithm computes an $S\in\Sscr$ with minimum $c(S)$ and adds a $T_S$-join $J$ with minimum $c(J)$.

\subsection{Best-of-Many Christofides}

Like \cite{AnKS12}, we begin by solving the natural LP relaxation: 

\begin{equation}
\label{stpathlp}
\hspace*{-0.4cm}
\begin{array}{lcrclcl}
\multicolumn{2}{l}{\min \ c(x)} && & & \\[0.5mm]
\mbox{subject to} && x(\delta(U)) &\ge& 2 && (\emptyset\not=U\subset V,\, |U\cap \{s,t\}| \text{ even}) \\
&& x(\delta(U)) &\ge& 1 && (\emptyset\not=U\subset V,\, |U\cap \{s,t\}| \text{ odd}) \\
&& x(\delta(v)) &=& 2 && (v\in V\setminus \{s,t\}) \\
&& x(\delta(v)) &=& 1 && (v\in \{s,t\}) \\
&& x_e & \ge & 0 && (e\in E)
\end{array}
\hspace*{-0.4cm}
\end{equation}
whose integral solutions are precisely the incidence vectors
of the edge sets of the Hamiltonian $s$-$t$-paths in $(V,E)$. 
This LP can be solved in polynomial time (either by the ellipsoid method (\cite{GroLS81})  or an extended formulation).
Let $x^*$ be an optimum solution. 
Then, $x^*$ (in fact every feasible solution) can be written as 
convex combination of spanning trees, i.e.\ as
$x^*=\sum_{S\in\Sscr}p_S \chi^S$,
where 
$p$ is a \emph{distribution} on $\Sscr$, 
i.e., $p_S\ge 0$ for all $S\in\Sscr$
and $\sum_{S\in\Sscr}p_S=1$.

By Carath\'eodory's theorem, we can assume that $p_S>0$ for less than $n^2$ spanning trees $(V,S)$.  
Such spanning trees and numbers $p_S$ can be computed in polynomial time,
using either the ellipsoid method or the splitting-off technique (cf.\ \cite{GenW15}).

The best-of-many Christofides algorithm does the following: 
 Compute an optimum solution $x^*$  for \eqref{stpathlp} and obtain a distribution $p$ with $x^*=\sum_{S\in\Sscr}p_S \chi^S$ as above. 
For each tree $(V, S)$ with $p_S > 0$, compute a minimum weight $T_S$-join $J_S$. 
Then, the multigraph $(V, S \dot{\cup} J_S)$ is an $\{s, t\}$-tour. We output the best of these.

We will fix $x^*$ henceforth. 
An important concept in \cite{AnKS12} and the subsequent works are the so-called {\em narrow cuts},
i.e., the cuts $C=\delta(U)$ with $x^*(C)<2$. We denote by $\Cscr$ the set of all narrow cuts. We are going to exploit their structure as well. 

\begin{lemma}[\cite{AnKS12}]
\label{chain}
The narrow cuts form a chain: there are 
sets $\{s\}=U_0\subset U_1\subset \cdots \subset U_{\ell-1}\subset U_\ell=V\setminus\{t\}$
so that $\Cscr=\{\delta(U_i):i=0,\ldots,\ell\}$.
These sets can be computed in polynomial time.
\end{lemma}
We number the narrow cuts $\Cscr=\{C_0,C_1\ldots,C_{\ell}\}$
with $C_i=\delta(U_i)$ ($i=0,\ldots,\ell$). 

\subsection{Gao trees \label{subsection:gao}}

Our work was inspired by the following idea of \cite{Gao13}:

\begin{theorem}[\cite{Gao13}]
\label{thmgao}
There exists a spanning tree $S\in\Sscr$ with $x^*_e>0$ for all $e\in S$ and $|C\cap S|=1$ for all $C\in\Cscr$.
\end{theorem}

In fact, \cite{Gao13} showed this for any vector $x \in \mathbb{R}^E_{\ge 0}$ with $x(\delta(U)) \ge 1$ for all $\emptyset \not = U \subset V$ and
$x(\delta(U)) \ge 2$ for all $\emptyset \not = U \subset V$ with $|U \cap \{s, t\}|$ even. 
For graph $s$-$t$-path TSP, one uses only variables corresponding to edges of the given graph. Then every spanning tree has cost $n - 1$. 
The approximation guarantee of $\frac{3}{2}$ then follows from the fact that for a tree $(V, S)$ with $|S\cap C| = 1$ for all $C \in \Cscr$,
 the vector $\frac{1}{2}x^*$ is in the $T_S$-join polyhedron.
But, as shown by \cite{Gao15}, for the general $s$-$t$-path TSP there may be no tree as in Theorem \ref{thmgao} whose cost is bounded by the LP value.  

Let us call a tree $S\in\Sscr$ a \emph{local Gao tree at $C$} if $|C\cap S|=1$. We call $S$ a \emph{global Gao tree} if it is a local Gao tree at every narrow cut. 

\cite{AnKS12} and \cite{Seb13} observed that for every distribution $p$ with $x^*=\sum_{S\in\Sscr} p_S \chi^S$ and every narrow cut $C\in\Cscr$,
at least a $2-x^*(C)$ fraction of the trees will be local Gao trees at $C$.
However, in general none of these trees will be a global Gao tree.

\subsection{Our contribution}

Our main contribution is a new structural result:
Starting from an arbitrary distribution of trees representing $x^*$, we can compute a new distribution in which a sufficient number of trees 
are Gao trees simultaneously for all sufficiently narrow cuts. 
For example, if $x^*(C)=\sfrac{3}{2}$ for all $C\in\Cscr$, at least half of our new distribution
will be made of global Gao trees.
Here is our main structure theorem:

\begin{theorem}\label{theorem:EnoughGoodTrees}
For every feasible solution $x^*$ of (\ref{stpathlp}),
there are $S_1,\ldots,S_r\in\Sscr$ and $p_1,\ldots,p_r>0$ with $\sum_{j=1}^r p_j=1$ such that
$x^* = \sum_{j=1}^r p_j \chi^{S_j}$ and for every $C\in\Cscr$ there exists a $k\in\{1,\ldots,r\}$ with 
$\sum_{j=1}^k p_j\ge 2-x^*(C)$ and $|C\cap S_j|=1$ for all $j=1,\ldots,k$.
\end{theorem}

Note that this result immediately implies Theorem \ref{thmgao}, simply by taking $S_1$.

However, we do not know whether such a distribution can be computed in polynomial time
(or even whether $r$ can be polynomially bounded). 
Therefore we will start with an arbitrary distribution $p$ and round the coefficients down to integral multiples of $\frac{\epsilon}{n^3}$ 
for a sufficiently small constant $\epsilon>0$. 
This way we will get a vector $x$ close to $x^*$ that we can write as
$x=\sum_{j=1}^r \frac{1}{r}\chi^{S_j}$, where $r\le \frac{n^3}{\epsilon}$ and $S_j\in\Sscr$ for $j=1,\ldots,r$.
We will work with $x$ in the following and consider the rounding error in Section \ref{sec:CorrectionVectors}.
We will show that $x\in\mathbb{R}_{\ge 0}^E$ satisfies the properties
\begin{equation}
\label{propertiesofx}
x(\delta(s))=x(\delta(t))=1 \qquad \text{ and } \qquad  x^*(F) - \epsilon \le x(F) \le x^*(F) + \epsilon \text{ for all } F\subseteq E.
\end{equation}

\begin{theorem}\label{theorem:EnoughGoodTreesRounded}
Given $S_1,\ldots,S_r\in\Sscr$, a feasible solution $x^*$ of (\ref{stpathlp}) and $\epsilon\ge 0$ such that
$x = \frac{1}{r} \sum_{j=1}^r \chi^{S_j}$ satisfies (\ref{propertiesofx}),
we can find $\hat{S}_1,\ldots,\hat{S}_r\in\Sscr$ in polynomial time such that
$x =\frac{1}{r} \sum_{j=1}^r \chi^{\hat{S}_j}$, 
and for every $C\in\Cscr$ there exists an $h\in\{1,\ldots,r\}$ with
$ \frac{h}{r} \ge 2 - x^*(C) - \epsilon$ and $|C\cap \hat{S}_j|=1$ for all $j=1,\ldots,h$.
\end{theorem}

Note that this theorem implies Theorem \ref{theorem:EnoughGoodTrees},
by letting $\epsilon=0$ and $x=x^* = \sum_{S \in \Sscr} p_S \chi^{S_j} = \sum_{j=1}^{r} \frac{1}{r} \chi^{S_j}$
 where all $p_S$ are integer multiples of $\frac{1}{r}$
for some (possibly exponentially large) natural number $r$ and taking $rp_S$ copies of each $S\in\Sscr$.

This paper is organized as follows: In Section \ref{sec:StructureTheorem}, we will prove Theorem \ref{theorem:EnoughGoodTreesRounded}. 
In Section \ref{sec:CorrectionVectors}, we will analyze the best-of-many Christofides algorithm on the resulting distribution.
Finally, in Section \ref{sec:Benefit} we provide the
computations needed to obtain the approximation guarantee of 1.566. 

\section{Proof of the Structure Theorem}\label{sec:StructureTheorem}

We will prove Theorem \ref{theorem:EnoughGoodTreesRounded} by starting with 
arbitrary trees $S_1,\ldots,S_r\in\Sscr$ with
$x = \frac{1}{r} \sum_{j=1}^r \chi^{S_j}$ satisfying (\ref{propertiesofx}), 
and successively exchange a pair of edges in two of the trees.
We will first satisfy the properties for the first tree $S_1$, then for $S_2$, and so on.
For each $S_j$, we will work on the narrow cuts $C_1,\ldots,C_{\ell-1}$ in this order; note that
$|C_0\cap S_j|=|C_{\ell}\cap S_j|=1$ always holds for all $j=1,\ldots,r$ due to the first property in \eqref{propertiesofx}.

In the following we write 
$$\theta_i:= \lceil r(2-x^*(C_i) - \epsilon) \rceil$$ 
for $i=0,\ldots,\ell$.
Note that $\theta_0  \ge \theta _i$ for all $i=1,\ldots,\ell$ because $x^*(C_i)\ge 1 = x^*(C_0) = x^*(\delta(s))$.
Our goal is to obtain $|S_j\cap C_i|=1$ whenever $j\le\theta_i$. 

Note that $|S_j\cap C_i|=1$ implies that $(V, S_j)[U_i]$ is connected, and we will
first obtain this weaker property by Lemma \ref{lemma:connectedLevelSet},
before obtaining $|S_j\cap C_i|=1$ by Lemma \ref{lemma:oneedgeincut}.

We need a few preparations. 

\begin{lemma}
\label{lemma:ChooseConnectedTree}
Let $S_1,\ldots,S_r\in\Sscr$ and $\epsilon\ge 0$ such that
$x = \frac{1}{r} \sum_{j'=1}^{r} \chi^{S_{j'}}$ satisfies (\ref{propertiesofx}).
Let $0\le h<i\le\ell$  and $1 \le j \le r$ with $j\le \theta_h$ and $j\le \theta_i$.
Let $M = U_i \setminus U_h$ or $M = U_i$. 
Then there exists an index $k\ge j$ such that 
$(V,S_k)[M]$ is connected.
\end{lemma}

\prove
Assume the above is not true. Then
$|E[M]\cap S_{j'}|\le |M|-1$ for all $j' < j$, and $|E[M]\cap S_{j'}|\le |M|-2$ for $j'\ge j$.
Therefore, 
\begin{align*}
x(E[M]) &= \sfrac{1}{r} \sum_{j' = 1} ^r |E[M]\cap S_{j'}| \\
&\le \sfrac{1}{r} \bigl((j - 1)(|M| - 1) + (r-j+1)(|M| - 2)\bigr) = |M| - 2 + \sfrac{j - 1}{r}.
\end{align*}

On the other hand, 
using (\ref{propertiesofx}) we have
$x(E[M]) + \epsilon \ge
x^*(E[M]) =
\frac{1}{2} \bigl( \sum_{v\in M} x^*(\delta(v)) - x^*(\delta(M)) \bigr)$.

If $M = U_i \setminus U_h$, this is equal to
$ |M| - \frac{1}{2} (x^*(C_{h}) + x^*(C_i)) + x^*(C_h\cap C_i) 
\ge |M| - \frac{1}{2} (x^*(C_{h}) + x^*(C_i))$.

If $M = U_i$,  this is equal to $ |M|  -\sfrac{1}{2} - \frac{1}{2}x^*(C_i) = |M|  -\sfrac{1}{2}(1 +x^*(C_i)) $.

Now,  $j\le \theta_i$ and $j\le \theta_h$ implies
$1 \le x^*(C_i) < 2 - \frac{j-1}{r} - \epsilon$ and $x^*(C_h) < 2 - \frac{j-1}{r} - \epsilon$.
Thus, 
$$x(E[M]) > -\epsilon + |M|  -\sfrac{1}{2}(2 - \frac{j-1}{r} - \epsilon + 2 - \frac{j-1}{r} - \epsilon)
= |M| - 2 + \sfrac{j-1}{r}, $$
a contradiction.
\endproof

\begin{lemma}
\label{lemma:ChooseSmallIntersection}
Let $S_1,\ldots,S_r\in\Sscr$ and $\epsilon\ge 0$ such that
$x = \frac{1}{r} \sum_{j'=1}^{r} \chi^{S_{j'}}$ satisfies (\ref{propertiesofx}).
Let $0\le h<i\le\ell$  and $1 \le j \le r$ with $j\le \theta_h$ and $j\le \theta_i$.
Then there exists an index $k\ge j$ such that 
 $S_k\cap C_h\cap C_i=\emptyset$.

\end{lemma}

\prove
Using first \eqref{propertiesofx} and then $x^*(C_{i'}) < 2 - \frac{j-1}{r} - \epsilon$ for $i' \in \{h, i\}$
and $x^*(U) \ge 2$ for $|U\cap \{s, t\}|$ even,   
we obtain
\begin{align*}
x(C_h\cap C_i)  - \epsilon &\le x^*(C_h\cap C_i) = \sfrac{1}{2}\bigl(x^*(C_h) + x^*(C_i) - x^*(\delta(U_i \setminus U_h))\bigr) \\
&< 2 - \sfrac{j-1}{r} - \epsilon  - 1 = \sfrac{r-j+1}{r} - \epsilon.
\end{align*}

Therefore, $\frac{1}{r} \sum_{j' = 1}^r |S_{j'} \cap C_h\cap C_i| =  x(C_h\cap C_i) < \frac{r-j + 1}{r}$,
 i.e.\  at most $r-j$ trees can contain an edge in $C_h\cap C_i$. 
\endproof

\begin{lemma}
\label{lemma:Choose1NarrowEdgeTree}
Let $S_1,\ldots,S_r\in\Sscr$ and $\epsilon\ge 0$ such that
$x = \frac{1}{r} \sum_{j'=1}^{r} \chi^{S_{j'}}$ satisfies (\ref{propertiesofx}). 
Let $1\le i\le \ell - 1$ and $j \le \theta_i$ with $|C_i \cap S_j| \geq 2$.
Then there exists an index $k > \theta_i$ with $|C_i \cap S_k|  = 1$. 
\end{lemma}

\prove
Suppose there exists no such $k$. Then we get $rx(C_i)=\sum_{j' = 1}^r |C_i\cap S_{j'}|  \geq (r - \theta_i + 1)2 + \theta_i - 1 
= 2r - \lceil  r(2-x^*(C_i)-\epsilon) \rceil  + 1 > 2r - r(2 - x^*(C_i) - \epsilon) = r(x^*(C_i)+\epsilon)$, 
which is a contradiction to (\ref{propertiesofx}). 
\endproof

\begin{lemma}
\label{lemma:increaseconnectivity}
Let $1\le i \le \ell-1$ and $M \subseteq U_i$  and $S_j,S_k\in\Sscr$ such that $(V,S_j)[M]$ is disconnected and $(V,S_k)[M]$ is connected
and $|S_j\cap \delta(U_i\setminus M)|\le 1$.
Then there exist edges $e\in S_j$ and $f\in S_k$ such that $S_j-e+f\in\Sscr$ and $S_k+e-f\in\Sscr$
and $e\notin E[U_i]$ and $f\in E[M]$.
\end{lemma}

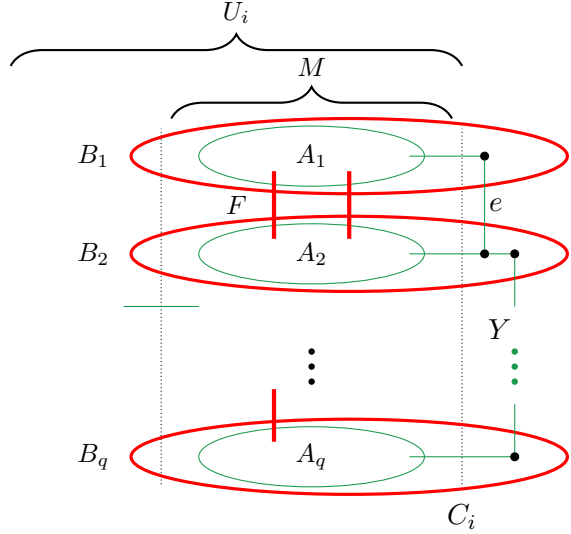
\begin{figure}[h]
\centering
\begin{tikzpicture}[]
\tikzstyle{vertex}=[circle, draw, fill, minimum width=3pt, inner sep = 0]

\foreach \x/\name\index in {{0/h/1}, {4/i/2}}
{
	\node (v\index ) at (\x, 0) {};
	\node (w\index ) at (\x, 5) {};		 
}
\draw[densely dotted] (v2) node[below]{$C_i$} -- (w2) ; 
\draw[densely dotted] (v1) -- (w1) ; 

\draw[thick, decorate,decoration={brace,amplitude=12pt}] (w1) -- (w2) node[midway, above,yshift=12pt,]{\small $M$};
\draw[thick, decorate,decoration={brace,amplitude=12pt}] (-2,5.7) -- (4,5.7) node[midway, above,yshift=12pt,]{\small $U_i$};

\draw [darkgreen] (-0.5, 2.5) -- (0.5, 2.5);

\foreach {\y} in {4.5, 3.2, 0.5}
{
	\draw[darkgreen](2, \y) ellipse (1.5cm and 0.4cm);
	\draw[red, very thick](2.5, \y) ellipse (2.9cm and 0.5cm);
}	
	\draw[darkgreen] (3.3, 4.5) --(4.3, 4.5);
	\draw[darkgreen] (3.3, 3.2) --(4.3, 3.2);
	\draw[darkgreen] (3.3, 0.5) --(4.7, 0.5);
	
\node at (2, 4.5){\textcolor{darkgreen}{\textcolor{black}{\small $A_{1}$}}};
\node at (2, 3.2){\textcolor{darkgreen}{\textcolor{black}{\small $A_{2}$}}};
\node at (2, 0.5){\textcolor{darkgreen}{\textcolor{black}{\small $A_{q}$}}};

\node at (-0.9, 4.5){\textcolor{red}{\textcolor{black}{\small $B_{1}$}}};
\node at (-0.9, 3.2){\textcolor{red}{\textcolor{black}{\small $B_{2}$}}};
\node at (-0.9, 0.5){\textcolor{red}{\textcolor{black}{\small $B_{q}$}}};

\draw[red, ultra thick] (1.5, 4.3) -- (1.5, 3.4);
\draw[red, ultra thick] (2.5, 4.3) -- (2.5, 3.4);
\draw[red, ultra thick] (1.5, 0.7) -- (1.5, 1.4);
\node at (1, 3.85){\textcolor{red}{\textcolor{black}{\small $F$}}};

\foreach \x/\y/\nr in {{4.3/4.5/1}, {4.7/3.2/2}, {4.3/3.2/3}, {4.7/0.5/4}}
	\node[vertex] (v\nr) at (\x, \y) {}; 
\foreach \i/\j in {{1/3}, {2/3}}
	\draw[darkgreen] (v\i) -- (v\j);
\draw[darkgreen] (v2) -- (4.7, 2.5);
\draw[darkgreen] (v4) -- (4.7, 1.2);
\foreach \y in {1.5, 1.7, 1.9}
{
	\draw [fill] (2, \y) circle (1 pt); 
	\draw [fill, darkgreen] (4.7, \y) circle (1 pt); 
}

\node at (4.5, 2.2){\textcolor{darkgreen}{\textcolor{black}{$Y$}}};
\node at (4.45, 3.85){\textcolor{darkgreen}{\textcolor{black}{$e$}}};
\node at (4.2, 3){\textcolor{darkgreen}{}};
 \end{tikzpicture} 
 \caption{Tree $S_j$ in green, $S_k$ in red/bold. In this example one can choose $p=1$.
 Note that $e$ could belong to $C_i$ (but not to $\delta(U_i \setminus M)$).}\label{fig:connectedLevelSet}
 \end{figure}

\prove
An illustration of the following proof can be found in Figure \ref{fig:connectedLevelSet}. 
Let $A_1,\ldots,A_q$ be the vertex sets of the connected components of $(V,S_j)[M]$; note that $q\ge 2$.

Let $F:=S_k\cap \bigcup_{p=1}^q (\delta(A_p)\setminus \delta(M))$ be the set of edges of $S_k$ between the sets $A_1,\ldots,A_q$.
Note that $F\subseteq E[M]$.
For $p=1,\ldots,q$ let $B_p$ be the set of vertices reachable from $A_p$ in $(V,S_k\setminus F)$.
Trivially, $A_p\subseteq B_p$ for all $p$, and $\{B_1,\ldots,B_q\}$ is a partition of $V$ because $(V,S_k)[M]$ is connected and $(V, S_k)$ is a tree. 
Let $Y$ be the union of the edge sets of the unique $v$-$w$-paths  in $S_j$ for all $v, w \in M$.
Note that $Y \subseteq E[ V \setminus (U_i \setminus M)]$ because $|S_j\cap \delta(U_i\setminus M)|\le 1$.

\medskip
{\bf Claim:} There exists an index $p\in\{1,\ldots,q\}$ and an edge $e\in Y\cap\delta(B_p)$ 
such that for every $p'\in\{1,\ldots,q\}\setminus \{p\}$ the path in $S_j$ from $A_p$ to $A_{p'}$ contains $e$.

\medskip
To prove the Claim, note that $(V,Y)$ consists of a tree and possibly isolated vertices. 
Choose an arbitrary root $z$ in this tree and take 
an edge $e \in Y\cap \bigcup_{p'= 1}^q \delta(B_{p'})$ with maximum distance from $z$. 

Let $D:=\{v \in V : e \text{ is on the $z$-$v$-path in } (V, Y)\}$. We will show that $D\cap M = A_p$ for some $p \in \{1, \ldots, q\}$. 
This will immediately imply that $p$ and $e$ satisfy the properties of the Claim.

Observe that $D \cap M \not = \emptyset$. Let $v \in D \cap M$ and $p$ such that $v \in A_p$.  
Since $(V,S_j)[A_p]$ is connected, this implies $A_p \subseteq D$. 
$(D, Y\cap E[D])$ is a tree and by the choice of $e$, it contains no edge from $\bigcup_{p'= 1}^q \delta(B_{p'})$. Therefore, $D \subseteq B_p$ and hence, $D\cap M = A_p$.  
The Claim is proved.

Now, take an index $p$ and an edge $e$ as in the Claim.
Consider the path $P$ in $S_k$ that connects the endpoints of $e$. Since $e \in \delta(B_p)$,
$P$ has an edge $f\in\delta(B_p)\cap S_k = \delta(A_p) \cap F$. Thus, $(V,S_k + e - f)$ is a tree.
The path in $S_j$ that connects the endpoints of $f$ contains $e$ by the Claim. 
Thus, $(V,S_j - e + f)$ is a tree. We have $f \in E[M]$ since $f \in F$, and $e \not \in E[U_i]$ as $e \in Y\cap \delta(B_p)$. 
\endproof

Now we are ready to prove our main lemmas:
  
\begin{lemma}\label{lemma:connectedLevelSet}
Let $S_1,\ldots,S_r\in\Sscr$ and $\epsilon\ge 0$ such that
$x = \frac{1}{r} \sum_{j'=1}^r \chi^{S_{j'}}$ satisfies (\ref{propertiesofx}).
Let $1\le i \le \ell - 1$ 
such that $j\le \theta_i$ and $|S_j\cap C_h|=1$ for all $h<i$ with $j\le \theta_h$.
Then we can find $\hat{S}_1,\ldots,\hat{S}_r\in\Sscr$ in polynomial time such that 
$x =\frac{1}{r} \sum_{j'=1}^{r} \chi^{\hat{S}_{j'}}$ and $\hat{S}_{j'}=S_{j'}$ for all $j'<j$
and $|\hat{S}_j\cap C_h|=1$ for all $h<i$ with $j\le \theta_h$, 
and $(V,\hat{S}_j)[U_i]$ is connected.
\end{lemma}

\prove
Assume that $(V,S_j)[U_i]$ is disconnected, i.e.,
$|S_j\cap E[U_i]|<|U_i|-1$.
Let $h$ be the largest index smaller than $i$ with $j\le \theta_h$. 
Such an index must exist because $\theta_0\ge\theta_i\ge j$. 

\medskip
{\bf Case 1:} $|S_j\cap E[U_i\setminus U_h]| < |U_i\setminus U_h|-1$.

Let $M:=U_i\setminus U_h$.
Note that $|S_j\cap \delta(U_i\setminus M)| = |S_j\cap C_h|=1$.
Since $(V, S_j)[M]$ is not connected, by Lemma \ref{lemma:ChooseConnectedTree} there exists an index $k > j$ such that $(V, S_k)[M]$ is connected.

Now we apply Lemma \ref{lemma:increaseconnectivity} and obtain two trees
$\hat{S}_j:=S_j-e+f$ and $\hat{S}_k:=S_k+e-f$ with $e\notin E[U_i]$ and $f\in E[M]$.

We have $|\hat{S}_j\cap E[U_i]|=|S_j\cap E[U_i]|+1$ and
$|\hat{S}_j\cap C_{h'}| \le |S_j\cap C_{h'}|$  for all $h'\le h$ and hence $|\hat{S}_j\cap C_{h'}| = 1$ for $h' \le h$ with $j \le \theta_{h'}$.
Note that $j>\theta_{i'}$ for all $h<i'<i$, so a new edge $f$ in such cuts $C_{i'}$ does no harm.

We replace $S_j$ and $S_k$ by $\hat{S}_j$ and $\hat{S}_k$ and leave the other trees unchanged.
If $(V,\hat{S}_j)[U_i]$ is still not connected, we iterate.

\medskip
{\bf Case 2:} $|S_j\cap E[U_i\setminus U_h]| = |U_i\setminus U_h|-1$.

\begin{figure}[h]
\centering
\tikzstyle{vertex}=[circle, draw, fill, minimum width=3pt, inner sep = 0]
\begin{tikzpicture}[]

\draw[densely dotted] (3, 0.5) -- (3, 4.5);
\node[anchor = north] at (3, 0.5) {{\small $C_i$}};
\draw[densely dotted] (0, 0.5) -- (0, 4.5);
\node[anchor = north] at (0, 0.5) {{\small $C_h$}};

\foreach \x/\y/\nr in {{-1/3.5/0}, {4/3.5/1}, {3.6/2/2}, {2/2/3}, {1/2.5/4}, {-0.7/2.5/5}, {1.5/1/8}}
{
	\node[vertex] (v\nr) at  (\x, \y){};
}
	\node [anchor = south] at (v0){\small $v$};
	\node [anchor = south] at (v1){\small $w$};

\draw [darkgreen] (v2)-- (v3);
\draw [darkgreen] (v0)-- (v1) node[midway, below]{\textcolor{black}{\small $\hat{e}$}};
\draw[dashed, darkgreen] (v4) -- (v5) node[midway, below]{\textcolor{black}{\small $f$}};
\draw[rounded corners, darkgreen] (3.3, 2.6) -- (3.3, 4) --(5, 4) --(5, 2.6) -- cycle;
\node at (5.3, 3.5) {\textcolor{darkgreen}{\textcolor{black}{\small $W$}}};
\draw[bend left = 90 , red, dashed, ultra thick] (v1) to (v8);
\draw[bend left = 90 , red, dashed, ultra thick] (v8) to (v0);

\draw[thick, decorate,decoration={brace,amplitude=12pt}] (0, 3.6) -- (3, 3.6) node[midway, above,yshift=12pt,]{\scriptsize connected in $S_j$};

\draw[thick, decorate,decoration={brace,amplitude=12pt}] (-1.5, 4.4) -- (3, 4.4) node[midway, above,yshift=12pt,]{\scriptsize  $M$};
 \end{tikzpicture} 
 \caption{Tree $S_j$ in green, $\hat{S}_{k'}$ in red/bold. 
 The dashed edge $f$ is added to $S_j$ by applying Lemma \ref{lemma:increaseconnectivity}.}\label{fig:connectedReconfiguration}
 \end{figure}
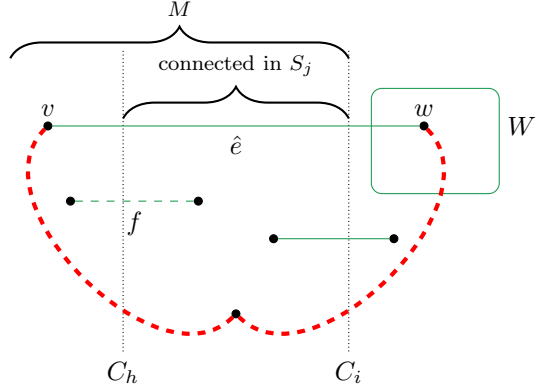

$(V,S_j)[U_h]$ is connected since $|S_j\cap C_h|=1$.
Moreover, $(V, S_j)[U_i\setminus U_h]$ is connected, but $(V, S_j)[U_i]$ is disconnected. 
Therefore, $S_j$ must contain an edge in $C_i \cap C_h$ and $S_j\cap C_h\subset S_j\cap C_i$ 
and $(V,S_j)[U_i]$ has exactly two connected components:
$U_h$ and $U_i\setminus U_h$. We will illustrate the following in Figure \ref{fig:connectedReconfiguration}.

{\bf Case 2a:} $h>0$.
Let  $g$ be the largest index smaller than $h$ with $j\le \theta_g$.
Set $M:=U_i\setminus U_g$. 
Note that $|S_j\cap \delta(U_i\setminus M)| = |S_j\cap C_g|=1$.
By Lemma \ref{lemma:ChooseConnectedTree} there exists an index $k \ge j$ with $(V, S_k)[M]$ connected. 
As $(V, S_j)[M]$ is not connected, we have $k > j$.

{\bf Case 2b:} $h=0$.
Set $M:=U_i$.
Note that $|S_j\cap \delta(U_i\setminus M)| = |S_j\cap \delta(\emptyset)|=0$.
By Lemma \ref{lemma:ChooseConnectedTree}, there exists an index $k \ge j$ such that $(V, S_k)[M]$ is connected.
Since $(V, S_j)[M]$ is not connected, $k > j$. 

Note that in both cases 2a and 2b, $(V,S_j)[M]$ is disconnected.
Now we apply Lemma \ref{lemma:increaseconnectivity} and obtain two trees
$\hat{S}_j:=S_j-e+f$ and $\hat{S}_k:=S_k+e-f$ with $e\notin E[U_i]$ and $f\in E[M]$.
We replace $S_j$ and $S_k$ by $\hat{S}_j$ and $\hat{S}_k$ and leave the other trees unchanged.
Then $(V,\hat{S_j})[U_i]$ is connected.
We have $|\hat{S}_j\cap C_{h'}|=1$ for all $h'<h$ with $j \le \theta_{h'}$, but we may have $|\hat{S}_j\cap C_h|=2$.

Assume $|\hat{S}_j\cap C_h|=2$ (otherwise we are done). 
Then $\hat{S}_j\cap C_h\cap C_i = S_j\cap C_h\cap C_i = S_j \cap C_h$,
and this set contains precisely one edge $\hat{e}=\{v,w\}$ (where $v\in U_h$ and $w\in V\setminus U_i$).
By Lemma \ref{lemma:ChooseSmallIntersection} there exists an index $k' > j$ with
 $\hat{S}_{k'}\cap C_h\cap C_i = \emptyset$.
 
Let $W$ be the set of vertices reachable from $w$ in $(V,\hat{S}_j\setminus C_i)$.
Since $(V,\hat{S}_j)[U_i]$ is connected, $\hat{S}_j\cap\delta(W)=\{\hat{e}\}$.
The unique path in $(V,\hat{S}_{k'})$ from $v$ to $w$ contains at least one edge $\hat{f}\in\delta(W)$.
Note that $\hat{f}\notin C_h$ by the choice of $\hat{S}_{k'}$. We replace $\hat{S}_j$ and $\hat{S}_{k'}$ by 
$\hat{\hat{S}}_j:=\hat{S}_j-\hat{e}+\hat{f}$ and $\hat{\hat{S}}_{k'}:=\hat{S}_{k'}+\hat{e}-\hat{f}$.
Then $(V,\hat{\hat{S}}_j)[U_i]$ is still connected and $|\hat{\hat{S}}_j\cap C_{h'}|=1$ for all $h'<i$ with $j\le \theta_{h'}$.
\endproof

\begin{lemma}\label{lemma:oneedgeincut}
Let $S_1,\ldots,S_r\in\Sscr$ and $\epsilon\ge 0$ such that
 $x=\frac{1}{r} \sum_{j'=1}^{r} \chi^{S_{j'}}$ satisfies (\ref{propertiesofx}).
Let $1 \le i \le \ell-1$ and 
$j\le \theta_i$ such that $(V,S_j)[U_i]$ is connected
and $|S_j\cap C_h|=1$ for all $h < i$ with $j\le \theta_h$.
Then we can find $\hat{S}_1,\ldots,\hat{S}_r\in\Sscr$ in polynomial time such that 
$x=\frac{1}{r} \sum_{j'=1}^{r} \chi^{\hat{S}_{j'}}$ and
$\hat{S}_{j'}=S_{j'}$ for all $j'<j$ and
$|\hat{S}_j\cap C_h|=1$ for all $h\le i$ with $j\le \theta_h$.
\end{lemma}

\begin{figure}[h]
\centering
\tikzstyle{vertex}=[circle, draw, fill, minimum width=3pt, inner sep = 0]
\begin{tikzpicture}[]

\draw[densely dotted] (0, 0) -- (0, 4.5);
\node[anchor = north] at (0, 0) {{\scriptsize $C_i$}};

\foreach \y/\n/\na/\nr in {{3.5/v/w/1}, {2/ / /2}, {1/x/y/3}}
{
	\node[vertex] (v\nr) at  (-0.5, \y){};
	\node[vertex] (w\nr) at  (0.5, \y){};
	\draw [darkgreen] (v\nr)-- (w\nr);
	\node [anchor = south] at (v\nr){\small $\n$};
	\node [anchor = south] at (w\nr){\small $\na$};
}
\node [anchor = south] at (0.15, 3.15){\textcolor{darkgreen}{\textcolor{black}{\small $e$}}};
\draw [red, ultra thick] (v3)-- (w3);

\draw[rounded corners, darkgreen] (0.3, 0.5) -- (0.3, 2.5) --(1.3, 2.5) --(1.3, 0.5) -- cycle;
\node at (0.65, 0.3) {\textcolor{darkgreen}{\textcolor{black}{\small $A$}}};

\draw[rounded corners, darkgreen] (0.3, 3) -- (0.3, 4) --(1.3, 4) --(1.3, 3) -- cycle;
\node at (0.65, 4.3) {\textcolor{darkgreen}{\textcolor{black}{\small $B$}}};

\draw [bend right = 40, darkgreen, dashed] (w3) to (w2);

\node[vertex] (w4) at (1, 3.5){};
\node[vertex] (w5) at (2, 3.5){};

\draw [red, dashed, ultra thick ] (w1)-- (w4);
\draw [red, ultra thick ] (w4)-- (w5) node[midway, above]{\textcolor{black}{\small $f$}};
\draw [red, dashed, ultra thick, bend left = 30] (w5)  to (w3);
\node[ anchor = west] at (1.8,2.5) {\textcolor{red}{\textcolor{black}{\small $P$}}};
 \end{tikzpicture} 
 \caption{Tree $S_j$ in green, $S_k$ in red/bold}\label{fig:oneedgeincut}
 \end{figure}
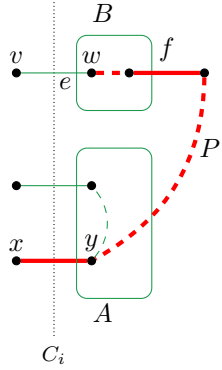

\prove
Suppose $|S_j\cap C_i|\ge 2$.
Then by Lemma \ref{lemma:Choose1NarrowEdgeTree} there exists an index $k>\theta_i$ with  $|S_k\cap C_i|=1$.
We will swap a pair of edges, reducing $|S_j\cap C_i|$ and increasing $|S_k\cap C_i|$ while maintaining the other properties.
An illustration of the following construction is given in Figure \ref{fig:oneedgeincut}.

Let $S_k\cap C_i=\{\{x,y\}\}$ with $x\in U_i$ and $y\in V\setminus U_i$.
Let $A$ be the set of vertices reachable from $y$ in $(V, S_j\setminus C_i)$. Note that $A\cap U_i=\emptyset$.
We have $|\delta(A)\cap S_j\cap C_i|=1$ because $(V,S_j)[U_i]$ is connected.
So let $e=\{v,w\}\in (S_j\cap C_i)\setminus\delta(A)$, with $v\in U_i$ and $w\in V\setminus U_i$.
Let $B$ be the set of vertices reachable from $w$ in $(V, S_j\setminus C_i)$.
We have $w\in B$, $y\in A$, and $A\cap B=\emptyset$ by the choice of $e$.
Consider the path $P$ in $S_k$ from $w$ to $y$.
Note that $P$ does not contain any vertex in $U_i$ because $|C_i\cap S_k|=1$.
But $P$ contains at least one edge $f\in\delta(B)$. 

We will swap $e$ and $f$. 
Since $S_k\cap C_i = \{\{x, y\}\}$, the $w$-$v$-path in $S_k$ contains $P$. 
Therefore, $\hat{S}_k := S_k + e - f$ is a tree. On the other hand, the path in $S_j$ connecting the endpoints of $f$ must use an edge in $\delta(B)$. 
Since $S_j\cap E[U_i]$ is connected and $S_j\cap(\delta(B)\setminus C_i)=\emptyset$, 
$e$ is the only edge in $\delta(B) \cap S_j$ and thus, $\hat{S}_j := S_j + f - e$ is a tree.

Since $f \in E[V\setminus U_i]$ and $e \in C_i$, we have $|\hat{S}_j\cap C_i|=|S_j\cap C_i|-1$ and $|\hat{S}_j\cap C_{h}|=|S_j\cap C_{h}|$ for all $h < i$ with $j \le \theta_{h}$.
Moreover, $(V,\hat{S}_j)[U_i]$ is still connected.
As before, we replace $S_j$ and $S_k$ by $\hat{S}_j$ and $\hat{S}_k$ and leave the other trees unchanged.
 If $|\hat{S}_j \cap C_i| > 1$, we iterate.
\endproof

Now the proof of Theorem \ref{theorem:EnoughGoodTreesRounded} is a simple induction. 
We scan the indices of the trees $j=1,\ldots,r$ in this order.
For each $j$, we consider all narrow cuts $C_i$ with $j\le\theta_i$. 
Since $x$ satisfies \eqref{propertiesofx}, 
we always have $|S_j\cap C_0|=1$ and $|S_j\cap C_{\ell}|=1$ for all $j=1,\ldots,r$.
Now let $i\in\{1,\ldots,\ell-1\}$ with $j\le\theta_i$.
Assuming $|S_j\cap C_h|=1$ for all $h<i$ with $j\le\theta_h$,
we first apply 
Lemma \ref{lemma:connectedLevelSet} 
and then Lemma \ref{lemma:oneedgeincut}.
The new tree then satisfies $|S_j\cap C_h|=1$ for all $h\le i$ with $j\le\theta_h$, and $S_1, \ldots S_{j - 1}$ remain unchanged.
 
 \section{Analysis of the Approximation Ratio}\label{sec:CorrectionVectors}

In this section, we will explain and analyze the algorithm that was sketched in Section \ref{sec:Introduction}. 
As mentioned before, we use the best-of-many Christofides algorithm on a special distribution of trees. 
To obtain these trees, we start with an optimal solution $x^*$ of the LP \eqref{stpathlp} 
and a distribution $p$ 
with $x^* = \sum_{S \in \Sscr}p_S\chi^S$ and round down the coefficients to integer multiples
of $\frac{\epsilon}{n^3}$. By taking multiple copies of trees, we will obtain trees $S_1, \ldots, S_r$ on which we can apply Theorem \ref{theorem:EnoughGoodTreesRounded}. 
Now, there is a small contribution of trees that is lost due to the rounding error. 
We will deal with these separately in the analysis and use that they can only represent a small fraction of $x^*$. 
Best-of-many Christofides will then be applied to the union of the trees constructed by Theorem \ref{theorem:EnoughGoodTreesRounded} (we will call the set of these trees $\hat{\Sscr}$) and the "leftover" trees. 
How exactly we deal with rounding will be explained in the proof of Theorem \ref{theorem:ApproxGuarantee}. 

Our goal is to choose for each tree $S$ a cheap vector $y^S$ in the $T_S$-join polyhedron 
to bound the cost of parity correction for $S$. Clearly, $y^S = x^*$ is a possible choice as 
$x^*(\delta(U)) \ge 1$ for all $\emptyset \not = U \subset V$. But for the trees in $\hat{\Sscr}$, we will construct better vectors. 

We follow the analysis of \cite{Vyg15} (based on \cite{AnKS12} and \cite{Seb13}). 
Every tree $S\in\Sscr$ can be partitioned into the edge set $I_S$ of the $s$-$t$-path in $S$
and $J_S$, which is a $T_S$-join.
For any narrow cut $C\in\Cscr$ let $e_C\in C$ be a minimum cost edge in $C$.
We will choose numbers $0 \le \beta < \frac{1}{2}$ and $0\le \gamma_{S}\le 1$  for all $S\in\Sscr$ later.
Then, 
for $S\in\Sscr$, we set 
\begin{equation}
\label{defrS}
z^S \ := \ \sum_{e\in I_S} (1-2\beta) \gamma_{S} \chi^{\{e\}}
+ \Csumeven \!\!\!\!\!
\max\bigl\{0, \left( \beta(2-x^*(C)-\epsilon) - (1-2\beta)\gamma_{S} \right)\bigr\} \chi^{\{e_C\}}.
\end{equation}
and
\begin{equation}
\label{correctionvector}
y^S \ := \ \beta(1+\epsilon) x^* + (1-2\beta)\chi^{J_S} +  z^S.
\end{equation}
Observe that $z^S(C) \ge \beta(2 - x^*(C) - \epsilon)$ for all $C\in \Cscr$ with $|S\cap C|$ even.
This implies that each $y^S$ is in the $T_S$-join polyhedron, because 
 $|\delta(U)\cap S| \text{ even} \Leftrightarrow   |U\cap T_S| \text{ odd}$ holds for all $U \subset V$ with $|U\cap \{s, t\}| = 1$. 
We adapt Definition 6 and Lemma 7 of \cite{Vyg15} as follows.

\begin{definition}
\label{defbenefit}
Given numbers $0\le \gamma_{S}\le 1$ for $S\in\Sscr$ and $\beta < \frac{1}{2}$,
we define the \emph{benefit} of $(S,C)\in\Sscr\times\Cscr$ to be
$b_{S,C}:=\min\left\{\frac{\beta (2-x^*(C)-\epsilon) }{1-2\beta}, \gamma_{S} \right\}$
if $|S\cap C|$ is even,
$b_{S,C}:=1-\gamma_{S}$ 
if $|S\cap C|=1$,
and $b_{S,C}=0$ otherwise.
\end{definition}

\begin{lemma}
\label{lemmaneedbenefit}
Let $0\le\beta<\frac{1}{2}$ and $0\le \gamma_{S}\le 1$ for $S\in\Sscr$. 
Let $p$ be an arbitrary distribution over $\Sscr$ and $x = \sum_{S\in\Sscr}p_S\chi^S$.
If
\begin{equation}
\label{enoughbenefit}
\sum_{S\in\Sscr} p_S b_{S,C} \ \ge \ \frac{\beta (2-x^*(C)-\epsilon) (\sum_{S \in \Sscr: |S\cap C| \text{even}} p_S)}{1-2\beta}
\end{equation}
for all $C\in\Cscr$,
then $\sum_{S\in\Sscr} p_S c(y^S) \le  (\beta + \epsilon\beta)c(x^*) + (1 - 2\beta)c(x)$.
\end{lemma}

\prove
First we note that 
$\sum_{S \in \Sscr}p_S c(z^S) \le (1 - 2\beta)\sum_{S \in \Sscr}p_Sc(I_S)$ 
is proved exactly as in the proof of Lemma 7 in \cite{Vyg15}.
The claim then follows from 

\begin{align*} \sum_{S \in \Sscr}p_Sc(y^S) = &\ \beta(1 + \epsilon)c(x^*) + (1 - 2\beta) \sum_{S \in \Sscr}p_Sc(J_S) + \sum_{S \in \Sscr}p_Sc(z^S)\\
\le&\ \beta(1 + \epsilon)c(x^*)  + (1 - 2\beta)\sum_{S\in\Sscr}p_S c(S). 
\end{align*} 
\mathendproof

We now show how to set the $\gamma$-constants in order to maximize the benefits, with the ultimate goal to choose $\beta$ as large as possible. 
\cite{Seb13} set all $\gamma_S$ to $\frac{1}{2}$. But we designed our $r$ trees in such a way that the trees with small indices
have only one edge in narrow cuts, thus trees that have an even number edges in a narrow cut tend to have larger indices. For example, for a narrow cut $C$ with $x(C)= \frac{3}{2}$,  trees with index $\le \frac{r}{2}$ have one edge in $C$ (thus it is better to choose a smaller $\gamma_S$ for these).
On the other hand, the trees that have an even number of edges in $C$ will have an index $> \frac{r}{2}$,
and for those it is better to choose a bigger $\gamma_S$.
Therefore we set $\gamma_S=\delta<\frac{1}{2}$ for the first half of the trees and $\gamma_S=1-\delta>\frac{1}{2}$ for the second half. More precisely:

\begin{lemma}
\label{lemmacomputebenefit}
Let $S_1,\ldots,S_r\in\Sscr$ and $\epsilon\ge 0$ such that
$x = \frac{1}{r} \sum_{j=1}^r \chi^{S_j}$ satisfies (\ref{propertiesofx}), $r$ is even,
and for every $C\in\Cscr$ there exists an $h\in\{1,\ldots,r\}$ with
$ \frac{h}{r} \ge 2 - x^*(C) - \epsilon$ and $|C\cap S_j|=1$ for all $j=1,\ldots,h$.

We set $\delta:=0.126$, and
$\gamma_{S_j} = \delta$ if $j \le \frac{r}{2}$ and $\gamma_{S_j} = 1-\delta$ otherwise.
Choose $\beta$ such that  $\frac{\beta}{1 - 2\beta} = 3.327$.
Then
\begin{equation}
\label{enoughbenefitrounded}
\sfrac{1}{r}\sum_{j=1}^r b_{S_j,C_i} \ \ge \ 3.327 \, (2-x^*(C_i)-\epsilon) \, \sfrac{1}{r}|\{j:|S_j\cap C_i| \text{ even}\}|
\end{equation}
for all $i=0,\ldots,\ell$.
\end{lemma}

As the proof is quite technical, we prove this lemma in Section \ref{sec:Benefit}.

Using Lemma \ref{lemmaneedbenefit} and \ref{lemmacomputebenefit}, 
we can show that the best-of-many Christofides algorithm run on the trees constructed in 
Theorem \ref{theorem:EnoughGoodTreesRounded} and the trees remaining after rounding 
is a $2 - \beta + \epsilon$-approximation for $\beta$ as in Lemma \ref{lemmaneedbenefit} and obtain: 

\begin{theorem}\label{theorem:ApproxGuarantee}
There is a $1.566$-approximation algorithm for the $s$-$t$-path TSP.
\end{theorem}

\prove
Let $x^*$ be an optimal solution for \eqref{stpathlp} and $p$ a distribution over $\Sscr$ with $x^* = \sum_{S \in \Sscr}p_S\chi^S$ such that $p_S > 0$ for at most $n^2$ trees $S$. 
As mentioned before, $x^*$ and $p$ can be found in polynomial time. 
Setting $p'_S:=\frac{\epsilon}{n^3}\lfloor \frac{n^3}{\epsilon}p_S\rfloor$
and $p''_S:=p_S-p'_S$ for $S\in\Sscr$, we have $\sum_{S\in\Sscr}p''_S\le \frac{\epsilon}{n}$ since  $|\{S \in \Sscr: p_S > 0\}| \le n^2$. 
By setting $r=\frac{2n^3}{\epsilon}\sum_{S\in\Sscr}p'_S$ (an even integer), 
we can obtain $r$ trees $S_1,\ldots,S_r$, such that each $S\in\Sscr$ appears $2\lfloor \frac{n^3}{\epsilon}p_S\rfloor$ times in this list.

Let $x = \frac{1}{r}\sum_{j=1}^r \chi^{S_j} $. We show that $x$ fulfills property \eqref{propertiesofx}.  
Since all trees $(V, S)$ with $p_S > 0$ have $|\delta(s) \cap S| = |\delta(t)\cap S| = 1$, we have $x(\delta(s)) = x(\delta(t)) = 1$. 
It remains to show $x^*(F) - \epsilon \le x(F) \le x^*(F) + \epsilon$ for all $F \subseteq E$:
\begin{align*}
x(F) - x^*(F) =& \sum_{S \in \Sscr}\left(\sfrac{p_S'}{\sum_{S \in \Sscr} p_S'} - p_S \right)|S \cap F| \\
 =& \sum_{S \in \Sscr}\frac{p_S - p_S'' - (1 - \sum_{S \in \Sscr} p_S'')p_S}{1 - \sum_{S \in \Sscr} p_S''}|S \cap F| \\
 =& \sum_{S \in \Sscr}\Bigl(-p_S'' + p_S{\textstyle\sum_{S \in \Sscr} p_S''}  \Bigr)\frac{|S \cap F|}{1 - \sum_{S \in \Sscr}p_S''}
 \end{align*}

Since $|S \cap F| \le |S| = n-1$ for all $S \in \Sscr$, we have 
$x(F)-x^*(F)\le (\sum_{S \in \Sscr}p_S'') \bigl( \sum_{S \in \Sscr}(p_S\frac{|S \cap F|}{1 - \sum_{S \in \Sscr}p_S''}) \bigr) 
\le \frac{\frac{\epsilon}{n}(n - 1)}{1 - \frac{\epsilon}{n}} \le \epsilon$.

Similarly, $x(F)-x^*(F)\ge -\sum_{S \in \Sscr}p_S'' \frac{|S \cap F|}{1 - \sum_{S \in \Sscr}p_S''} 
\ge -\frac{\frac{\epsilon}{n}(n-1)}{1 - \frac{\epsilon}{n}} \ge -\epsilon$.

This shows \eqref{propertiesofx}.

Using Theorem \ref{theorem:EnoughGoodTreesRounded} we transform this list to 
$\hat{S}_1,\ldots,\hat{S}_r$ fulfilling the statement of the theorem.
Therefore, we can apply Lemma \ref{lemmacomputebenefit} to $\hat{S}_1, \ldots, \hat{S}_r$ and $x$ 
and obtain inequality \eqref{enoughbenefit} for $\frac{\beta}{1 - 2\beta} = 3.327$. Thus, we will be able to apply Lemma \ref{lemmaneedbenefit} in the analysis. 

Our algorithm does parity correction for each of $\hat{S}_1, \ldots, \hat{S}_r$ and for all trees $S$ with $p_S'' > 0$ and returns the best resulting tour. 
We obtain a bound on the value of the best output by an averaging argument. 
For that, we use $y^{\hat{S}_j}$ as defined above in \eqref{correctionvector} 
to bound the parity correction cost for trees $\hat{S}_j$ and $x^*$ for other trees. 

We have $x = \frac{1}{r}\sum_{j=1}^r \chi^{\hat{S}_j} 
= \frac{1}{r}\sum_{j=1}^r \chi^{S_j} 
= \frac{1}{\sum_{S\in\Sscr}p'_S}\sum_{S\in\Sscr}p'_S\chi^S$.
Therefore,
\begin{align}\label{xinequality}
c(x) = \frac{1}{\sum_{S\in\Sscr}p'_S}\sum_{S\in\Sscr}p'_S c(S) \le \frac{1}{\sum_{S\in\Sscr}p'_S}\sum_{S\in\Sscr}p_S c(S) =  \frac{1}{\sum_{S\in\Sscr}p'_S}c(x^*)
.\end{align}


Then, using Lemma \ref{lemmaneedbenefit} in $(*)$ and \eqref{xinequality} in $(**)$,
\begin{align*}
&\min \left\{ \min_{j=1}^r \bigl(c(\hat{S}_j)+c(y^{\hat{S}_j})\bigr), \min_{S \in \Sscr: p''_S>0} \bigl(c(S) + c(x^*)\bigr) \right\} \\
&\le  \sum_{S\in\Sscr}p'_S \sum_{j=1}^r \sfrac{1}{r} \bigl(c(\hat{S}_j) + c(y^{\hat{S}_j})\bigr) + \sum_{S \in \Sscr}p''_S \bigl(c(S) + c(x^*) \bigr) 	\\
&= \sum_{S \in \Sscr}p'_S c(S) +  \sum_{S\in\Sscr}p'_S \sum_{j=1}^r \sfrac{1}{r} c(y^{S_j}) +  \sum_{S \in \Sscr}p''_S c(S) + \sum_{S \in \Sscr}p''_S c(x^*) \\
&= \Bigl(1+ \sum_{S \in \Sscr}p''_S \Bigr) c(x^*) + \sum_{S \in \Sscr}p'_S \sum_{j=1}^r \sfrac{1}{r} c(y^{S_j}) 	\\
&\overset{(*)}{\le} \Bigl(1+ \sum_{S \in \Sscr}p''_S \Bigr) c(x^*) + \sum_{S \in \Sscr}p'_S \bigl((\beta+\epsilon\beta)c(x^*) + (1 - 2\beta)c(x)\bigr) \\
&\overset{(**)}{\le} \Bigl(1+ \sum_{S \in \Sscr}p''_S \Bigr) c(x^*) + \sum_{S \in \Sscr}p'_S (\beta+\epsilon\beta)c(x^*) + (1 - 2\beta)c(x^*)\\
&\le \Bigl(2-\beta +\epsilon\beta + \sum_{S \in \Sscr}p''_S \Bigr) c(x^*)  \\
&< (2 - \beta + \epsilon)c(x^*).
\end{align*}

In the last inequality, we used $ \sum_{S \in \Sscr}p''_S \le \frac{\epsilon}{n}$.
Moreover, $\frac{\beta}{1 - 2\beta} =  3.327$ is equivalent to $\beta = \frac{3.327}{7.654} > 0.434 + 0.0006$. 

Thus, for $\epsilon \le 0.0006$, our approximation ratio is at most $2 - 0.434 = 1.566$. 
\endproof

\section{Computing the Benefit}\label{sec:Benefit}

In this section we prove Lemma \ref{lemmacomputebenefit}.

Fix $i\in\{0,\ldots,\ell\}$.
We write $\xi:=x^*(C_i)+\epsilon$, $\pi:=\frac{1}{r}|\{j:|S_j\cap C_i| \text{ even}\}|$, and $\rho:=\frac{1}{r}\theta_i -(2-x^ *(C_i)-\epsilon)$.
That means 
$\rho r$ is the amount of  $\theta_i$ due to rounding up (remember that $\theta_i = \lceil r(2 - x^*(C_i) - \epsilon) \rceil$).
We have $\pi \le x(C_i)-1 \le x^*(C_i)+\epsilon - 1 = \xi - 1$: 
The first inequality follows from $x(C_i) \ge \frac{1}{r}(|\{S_j: |C_i \cap S_j| \ge 2\}|  + r) \ge \pi + 1$, the second is due to (\ref{propertiesofx}).

Clearly, \eqref{enoughbenefitrounded} holds if $\pi = 0$ or $\xi \ge 2$, thus we assume this is not true. 

Suppose $3.327(2-\xi)\le \delta$.
Consequently, $b_{S_j, C_i} = \frac{\beta (2-\xi) }{1-2\beta} = 3.327(2-\xi)$ for all $j$ with $|S_j \cap C_i|$ even.
Thus,  \eqref{enoughbenefitrounded} trivially holds and we assume from now on $3.327(2-\xi) > \delta$. 
For convenience of notation, we define $\mu = \min\{1-\delta, 3.327(2-\xi)\}$.

First note that among the trees $S_j$ with $j > \theta_i$, we either have trees with $|S_j\cap C_i|$ even or additional trees with $|S_j\cap C_i| = 1$, 
which will guarantee us a certain amount of benefit:
If there is an $f$ fraction of trees $S_j$ with $|S_j\cap C_i|\ge 3$, then
$3f+(1-f)+\pi \le x(C_i) \le x^*(C_i)+\epsilon = \xi$, so $f\le \frac{\xi-\pi-1}{2}$. 

Let us distinguish several cases. 

\bigskip
{\bf Case 1:} $\xi\le\frac{3}{2}$. Then
$\theta_i\ge\frac{r}{2}$, and the first half of the trees have only one edge in $C_i$. 

We have $3.327(2 - \xi) \ge 3.327\cdot \frac{1}{2} > 1 -  \delta$. Therefore, $\mu=1-\delta$.  
Considering all trees, we obtain the following lower bounds for the benefit:

\begin{center}
\renewcommand\arraystretch{1.2}
\begin{tabular}{l|c|c|c}
$j$ & $|S_j\cap C_i|$ & fraction of trees & $b_{S_j,C_i}$ \\\hline
$j \le \frac{r}{2}$                   & 1 & $\frac{1}{2}$ & $1-\delta$ \\
$\frac{r}{2} < j \le \theta_i$ & 1 & $ \frac{\theta_i}{r} - \frac{1}{2} = \frac{3}{2}-\xi+\rho$ & $\delta$ \\
$\theta_i<j$                         & 2,4,... & $\pi$ & $\mu$ \\
$\theta_i<j$                         & 3,5,... & $\le \frac{1}{2}(\xi-\pi-1)$ & 0 \\
$\theta_i<j$                         & 1 & $\ge \frac{1}{2}(\xi-\pi-1)-\rho$ & $\delta$ \\
\hline
\end{tabular}
\end{center}

\begin{figure}
\centering
\begin{tikzpicture}[scale  = 1]

\draw [thick] (0,2) node (yaxis) [left] {}  |- (4,0) node (xaxis) [below] {};
\draw[thick] (2, -0.2) node (mid) [below] {\small $\sfrac{1}{2}$} -- (2, 0);
\draw[thick] (4, -0.2) node (mid) [below] {\small $1$} -- (4, 0);
\draw[thick, fill = grey] (0, 0.4) rectangle (2, 2);

\draw[thick] (-0.2, 0.4) node (del) [left] {\small $\delta$} -- (0, 0.4);
\draw[thick] (-0.2, 1.6) node (del) [left] {\small $1-\delta$} -- (0, 1.6);
\draw[thick] (-0.2, 2) node (del) [left] {\small $1$} -- (0, 2);
\draw[thick] (-0.2, 0) node (del) [left] {\small $0$} -- (0, 0);
\draw[thick] (0, -0.2) node (del) [below] {\small $0$} -- (0, 0);

\draw[thick, fill = grey] (2, 1.6) rectangle (2.4, 2);
\draw[thick] (2.4, -0.2) node (mid) [below] {\small $\sfrac{\theta_i}{r}$} -- (2.4, 0);

\draw[thick, pattern = dots, pattern color = darkgreen] (2.4, 0) rectangle (3, 1.6);

\draw[thick,decorate,decoration={brace,amplitude=5pt}] (3, 0) -- (2.4, 0) node[midway, below,yshift=-3pt,]{\small $\pi$};

\draw[thick, pattern = north east lines, pattern color = blue] (3.5, 1.6) rectangle (4, 2);
        
\end{tikzpicture}
\caption{On the $x$-axis, we have the fraction of trees, on the y-axis their benefit. A lower bound for the benefit for trees $S_j$ is given by the grey area for $j \le \theta_i$, the green (dotted) area for $|S_j\cap C_i|$ even and the blue (lined) area otherwise.}\label{fig:benefitarea}
\end{figure}
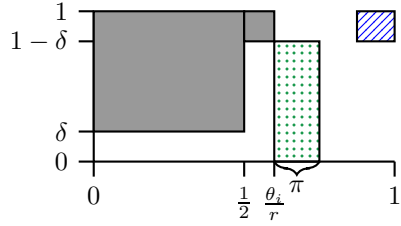

Thus, we can bound the weighted benefit (cf.\ Figure \ref{fig:benefitarea}): 
\begin{align*}
\sfrac{1}{r}\sum_{j = 1}^r b_{S_j, C_i}  
&\ge \sfrac{1-\delta}{2} + (\sfrac{3}{2} - \xi)\delta + \pi\mu + \sfrac{1}{2}(\xi-\pi-1)\delta \\
& = \sfrac{\delta + 1}{2} - \sfrac{\xi\delta}{2} + \pi(\mu - \sfrac{1}{2}\delta) \\
& = \sfrac{\delta + 1}{2} - \sfrac{\xi\delta}{2} + \pi(1 - \sfrac{3}{2}\delta) =: g_1(\xi, \pi)).
\end{align*}

To show \eqref{enoughbenefitrounded}, examine the minimum: 
\begin{align*}
\frac{\frac{1}{r}\sum_{j = 1}^r b_{S_j, C_i} }{ \pi(2-x^*(C_i)-\epsilon)} 
 \ge
&\min\left\{\frac{ g_1(\xi,\pi) }{ \pi(2-\xi)} : 1\le \xi \le \sfrac{3}{2}, \ 0< \pi\le \xi-1 \right\} \\
\overset{(*)}{ =}
&\min\left\{\frac{ \frac{\delta + 1}{2} - \frac{\xi\delta}{2} + (\xi-1)(1 - \frac{3}{2}\delta)}{ (\xi-1)(2-\xi)} : 1\le \xi \le \sfrac{3}{2} \right\} \\
\ge  
&\min\left\{\frac{2\delta - \frac{1}{2}  + \xi(1 - 2\delta)}{(\xi-1)(2-\xi)} : \xi \in \mathbb{R} \right\} 
 \ge 3.327.
\end{align*} 

The first inequality follows from the inequalities derived above.
The equality $(*)$ follows by considering the partial derivative of the function in $\pi$: 
$ \frac{\partial }{\partial \pi} \frac{g_1(\xi, \pi)}{\pi(2-\xi)}= \frac{1}{2-\xi}\big(-\frac{1}{\pi^2}\big(\frac{\delta + 1}{2} - \frac{\xi\delta}{2}\big)\big) $. 
This is negative for all $1\le \xi < 2$ and $\delta < 1$ and $\pi>0$.
Thus the minimum of $\frac{g_1(\xi, \pi)}{\pi(2-\xi)}$ is attained at a point where $\pi$ is maximal and it suffices to consider the case where $\pi = \xi - 1$. 

The last inequality is  matter of computation: We just have to verify that 
$2\delta - \frac{1}{2}  + \xi(1 - 2\delta) - 3.327(\xi-1)(2 - \xi) \geq 0$ for $\delta =  0.126$ and all $\xi\in\mathbb{R}$. 
The minimum is attained at $\xi \approx 1.39$. 
 
 \bigskip
{\bf Case 2:} $\xi>\frac{3}{2}$.
Then $\theta_i \le \frac{r}{2}$ because $r$ is even.
In this case, less than half of the trees are certain to have only one edge on $C_i$. 

Since $b_{S_j, C_i} = \mu > \delta$ for $j \ge \frac{r}{2}$ and $|S_j\cap C_i|$  even,
in the worst case those indices $j$ with $|S_j\cap C_i|$ even are as small as possible. Thus, we obtain the following lower bounds:

\begin{center}
\renewcommand\arraystretch{1.2}
\begin{tabular}{l|c|c|c}
$j$ & $|S_j\cap C_i|$ & fraction of trees & $b_{S_j,C_i}$ \\\hline
$j \le \theta_i$                   & 1 & $2 - \xi + \rho$ & $1-\delta$ \\
$ \theta_i < j \le \min\{\frac{r}{2}, \theta_i + r\pi\} $ & 2,4,... & $\min\{ \pi, -\frac{3}{2}  + \xi - \rho\}$& $\delta$ \\
$\frac{r}{2}<j \le \theta_i + r\pi$  & 2,4,... & $\max\{0, \frac{3}{2} -\xi +\rho + \pi \}$ & $\mu$ \\
$\theta_i<j$                         & 3,5,... & $\le \frac{1}{2}(\xi-\pi-1)$ & 0 \\
$\theta_i<j$                         & 1 & $\ge \frac{1}{2}(\xi-\pi-1)-\rho$ & $\delta$ \\\hline
\end{tabular}
\end{center}
\bigskip

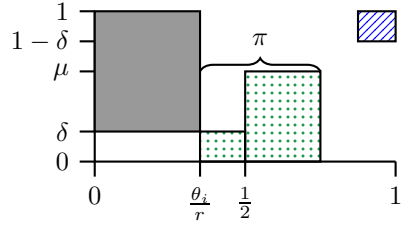
\begin{figure}
\centering
\begin{tikzpicture}[scale  = 1]

\draw [thick] (0,2) node (yaxis) [left] {}  |- (4,0) node (xaxis) [below] {};
\draw[thick] (2, -0.2) node (mid) [below] {\small $\sfrac{1}{2}$} -- (2, 0);
\draw[thick] (4, -0.2) node (mid) [below] {\small $1$} -- (4, 0);
\draw[thick, fill = grey] (0, 0.4) rectangle (1.4, 2);
\draw[thick] (1.4, -0.2) node (mid) [below] {\small $\sfrac{\theta_i}{r}$} -- (1.4, 0);

\draw[thick] (-0.2, 0.4) node (del) [left] {\small $\delta$} -- (0, 0.4);
\draw[thick] (-0.2, 1.2) node (del) [left] {\small $\mu$} -- (0, 1.2);
\draw[thick] (-0.2, 2) node (del) [left] {\small $1$} -- (0, 2);
\draw[thick] (-0.2, 0) node (del) [left] {\small $0$} -- (0, 0);
\draw[thick] (0, -0.2) node (del) [below] {\small $0$} -- (0, 0);

\draw[thick,  pattern = dots, pattern color = darkgreen] (1.4, 0) rectangle (2, 0.4);
\draw[thick, pattern = dots, pattern color = darkgreen] (2, 0) rectangle (3, 1.2);
\draw[thick,decorate,decoration={brace,amplitude=5pt}] (1.4, 1.2) -- (3, 1.2) node[midway, above,yshift=6pt,]{\small $\pi$};

\draw[thick,  pattern = north east lines, pattern color = blue] (3.5, 1.6) rectangle (4, 2);
\draw[thick] (-0.2, 1.6) node (del) [left] {\small $1 - \delta$} -- (0, 1.6);
        
\end{tikzpicture}
\caption{On the $x$-axis, we have the fraction of trees, on the y-axis their benefit. A lower bound for the benefit for trees $S_j$ is given by the grey area for $j \le \theta_i$, the green (dotted) area for $|S_j\cap C_i|$ even and the blue (lined) area for the remaining trees.}\label{fig:benefitarea2}
\end{figure}

Using the table we obtain (cf.\ Figure \ref{fig:benefitarea2}):
\begin{align*}
\sfrac{1}{r}\sum_{j = 1}^r b_{S_j, C_i} &\ge (2 - \xi + \rho)(1 - \delta) +  \min\{ \pi, -\sfrac{3}{2} + \xi - \rho\}\delta \\
& \quad +  \max\{0, \sfrac{3}{2} -\xi +\rho + \pi \}\mu +  (\frac{1}{2}(\xi-\pi-1) - \rho)\delta \\
&\ge (2 - \xi)(1 - \delta) +  \min\{ \pi, -\sfrac{3}{2}  + \xi\}\delta +  \max\{0, \sfrac{3}{2} -\xi + \pi \}\mu +  \sfrac{1}{2}(\xi-\pi-1)\delta \\
& \quad + \rho(1-2\delta)\\
&\ge (2 - \xi)(1 - \delta) +  \min\{ \pi, -\sfrac{3}{2}  + \xi\}\delta +  \max\{0, \sfrac{3}{2} -\xi + \pi \}\mu +  \sfrac{1}{2}(\xi-\pi-1)\delta \\
&=:g_2(\xi, \pi).
\end{align*}

Now, we distinguish two cases, depending on where the minimum in $g_2$ is attained. 

\bigskip
{\bf Case 2.1:} $\pi \le \xi - \frac{3}{2}$. 
Then $g_2(\xi, \pi) = (2 - \xi)(1 - \delta) + \pi\delta + \frac{\delta}{2}(\xi - \pi - 1)  = 2 - \frac{5}{2}\delta + (\frac{3}{2}\delta - 1)\xi + \frac{\delta}{2}\pi$.

As before, we minimize
\begin{align*}
\frac{\frac{1}{r}\sum_{j = 1}^r b_{S_j, C_i} }{ \pi(2-x^*(C_i)-\epsilon)} 
&\ge
\min\ \left\{
\frac{g_2(\xi, \pi)
}{ \pi(2-\xi)} : \sfrac{3}{2} < \xi <2 , \ 0< \pi \le \xi-\sfrac{3}{2} \right\} \\
& \overset{(*)}{ =}
\min \ \left\{ \frac{ 2 - \frac{5}{2}\delta + (\frac{3}{2}\delta - 1)\xi + \frac{\delta}{2}(\xi-\frac{3}{2})}{ (\xi-\frac{3}{2})(2-\xi)} : \sfrac{3}{2} < \xi <2 \right\} \\
&\ge \min\ \left\{ \frac{ 2 - \frac{13}{4}\delta + (2\delta - 1)\xi }{ (\xi-\frac{3}{2})(2-\xi)} : \xi\in\mathbb{R} \right\}
\ge 3.327
\end{align*} 

To show $(*)$, as before, we consider the partial derivative in $\pi$ and obtain
$ \frac{\partial }{\partial \pi} \frac{g_2(\xi, \pi)}{\pi(2-\xi)}= \frac{1}{2-\xi}\big(-\frac{1}{\pi^2}
\big(2 - \frac{5}{2}\delta + (\frac{3}{2}\delta - 1)\xi  \big)\big)$. 
Since $2 - \frac{5}{2}\delta + (\frac{3}{2}\delta - 1)\xi  = (2-\xi)(1 - \frac{3}{2}\delta) +\frac{\delta}{2}  > 0$,  
we have $ \frac{\partial }{\partial \pi} \frac{g_2(\xi, \pi)}{\pi(2-\xi)} < 0$ for all $\xi < 2$ and $\pi>0$, and the minimum is attained for $\pi =  \xi-\frac{3}{2}$. 
To verify the last inequality, checking $2 - \frac{13}{4}\delta + (2\delta - 1)\xi  - 3.327(\xi-\frac{3}{2})(2-\xi) \ge 0$  
for $\delta = 0.126$ and all $\xi\in\mathbb{R}$ is straightforward. 
(The minimum is attained at $\xi \approx 1.86$.)

\bigskip
{\bf Case 2.2:} $\pi > \xi - \frac{3}{2}$. 
Then $g_2(\xi, \pi) = (2 - \xi)(1 - \delta) + (\xi -\sfrac{3}{2})\delta +  (\sfrac{3}{2} -\xi + \pi)\mu +  \sfrac{1}{2}(\xi-\pi-1)\delta
 = 2 - 4\delta + \sfrac{3}{2}\mu + \xi(\sfrac{5}{2}\delta - 1 - \mu) + \pi(\mu-\frac{1}{2}\delta).$

Minimization yields
\begin{align*}
& \hspace*{0cm} \frac{\frac{1}{r}\sum_{j = 1}^r b_{S_j, C_i} }{ \pi(2-x^*(C_i)-\epsilon)} \\
&\ge 
\min\left\{\frac{g_2(\xi, \pi)}{ \pi(2-\xi)} :  \sfrac{3}{2} < \xi <2, \ \xi-\sfrac{3}{2} \le \pi \leq \xi - 1  \right\} \\
& \overset{(*)}{ =} \min\left\{\frac{2 - 4\delta + \sfrac{3}{2}\mu + \xi(\sfrac{5}{2}\delta - 1 - \mu) + \pi(\mu-\frac{1}{2}\delta)}{\pi(2-\xi)} : 
\sfrac{3}{2} < \xi <2, \ \pi  \in \{ \xi-\sfrac{3}{2}, \xi - 1\}  \right\} \\
& \overset{(**)}{\ge}   3.327.
\end{align*} 

To show $(*)$, again consider the partial derivative in $\pi$: 
$ \frac{\partial }{\partial \pi} \frac{g_2(\xi, \pi)}{\pi(2-\xi)}=
\frac{1}{2-\xi}\big(-\frac{1}{\pi^2} \big(2 - 4\delta + \sfrac{3}{2}\mu + \xi(\sfrac{5}{2}\delta - 1 - \mu)\big) \big )$.
Unlike the previous cases, this function can be positive or negative for $\delta = 0.126$ and $\frac{3}{2} < \xi < 2$ and $\pi>0$.
But for fixed $\xi$, it is either always  non-negative or always non-positive. 
Therefore the minimum is either attained for maximal or for minimal $\pi$, which explains the equality. 

We now show $(**)$.
Let us examine the two possible choices for $\pi$. 
For  $\pi = \xi-\frac{3}{2} $, we minimize $\frac{g_2(\xi, \xi - \frac{3}{2})}{( \xi - \frac{3}{2})(2 - \xi)}$, 
which is the same function as in Case 2.1 (note that the terms involving $\mu$ cancel).  

For $\pi = \xi-  1$, we get:
\begin{align}\label{minCase2.2}
 &\min\left\{\frac{2 - 4\delta + \sfrac{3}{2}\mu + \xi(\sfrac{5}{2}\delta - 1 - \mu) + (\xi - 1)(\mu-\frac{1}{2}\delta)}{ (\xi - 1)(2-\xi)} 
 :  \sfrac{3}{2} < \xi <2  \right\} \nonumber \\
\ge &\min\left\{\frac{2 - \frac{7}{2}\delta + \sfrac{1}{2}\mu + \xi(2\delta - 1) }{ (\xi - 1)(2-\xi)}: \xi \in \mathbb{R}  \right\}. 
\end{align} 

If $\mu = 1 - \delta$, \eqref{minCase2.2} equals  
 $\min\limits_{\xi \in \mathbb{R}}\left\{\frac{\frac{5}{2} - 4\delta  + \xi( 2\delta -1)}{ (\xi - 1)(2-\xi)} \right\}
= \min\limits_{\xi \in \mathbb{R}}\left\{\frac{2\delta -\frac{1}{2} + (3-\xi)( 1- 2\delta)}{ (\xi - 1)(2-\xi)} \right\}
= \min\limits_{\xi' \in \mathbb{R}}\left\{\frac{2\delta -\frac{1}{2} + \xi'( 1- 2\delta)}{ (\xi' - 1)(2-\xi')} \right\}$.

We replaced $\xi'=3-\xi$, which corresponds to mirroring at $\frac{3}{2}$.
This minimum was already shown to be at least $3.327$ in Case 1. 

If $\mu = 3.327(2 - \xi)$, then \eqref{minCase2.2} is equal to

 $$\min\left\{\frac{2 - \sfrac{7}{2}\delta + \xi(2\delta - 1) + \sfrac{1}{2}3.327(2 - \xi)}{ (\xi - 1)(2-\xi)}  :  \sfrac{3}{2} < \xi <2  \right\}.$$
 
 To see that this is at least $3.327$,
 it just remains to check that $2 - \sfrac{7}{2}\delta + \xi(2\delta - 1) + \sfrac{1}{2}3.327(2 - \xi) - 3.327(\xi - 1)(2-\xi) \geq 0$ for $\delta = 0.126$ and all $\xi \in \mathbb{R}$. 
 The minimum is attained at $\xi \approx 1.86$.

This completes the proof of Lemma \ref{lemmacomputebenefit}.

\section{Conclusion}

The approximation ratio can probably be improved slightly by choosing the $\gamma_{\hat{S}_j}$ differently,
but still depending only on $\frac{j}{r}$.
However, using an analysis based on Lemma \ref{lemmaneedbenefit},
one cannot obtain a better approximation ratio than $\frac{14}{9}$
because the benefit can never be more than one and there can be cuts $C$ with
$x^*(C)=\frac{3}{2}$ and $\sum_{S\in\Sscr:|S\cap C| \text{ even}}p_S =\frac{1}{2}$; therefore
$\frac{\beta}{1-2\beta}\le 4$.
Our ratio is already close to this threshold.

On the other hand, it is not impossible that the best-of-many Christofides algorithm
on a distribution like the one obtained in Theorem \ref{theorem:EnoughGoodTreesRounded}, or even on an arbitrary distribution,
has a better approximation ratio, maybe even $\frac{3}{2}$. 

%
%
%
%
%
%
%
%
%

{\small
\newcommand{\bib}[3]{\bibitem[\protect\citeauthoryear{#1}{#2}]{#3}}

}
\end{document}